\documentclass[aps,pra,twocolumn,superscriptaddress]{revtex4}
\usepackage{amssymb}
\usepackage{graphicx}
\usepackage{amsmath}
\usepackage{natbib}
\usepackage{bm}

\usepackage{color}

\def\ba{\begin{eqnarray}}
\def\ea{\end{eqnarray}}

\begin{document}

\title{Density Wave Superradiance of Photonic Fluid in Frustrated Triangle Lattice Cavity Arrays}

\author{Chao Feng}
\affiliation{Institute for Advanced Study, Tsinghua University, Beijing, 100084, China}
\author{Yu Chen}
\email{yuchen.physics@cnu.edu.cn}
\affiliation{Department of Physics, Capital Normal University, Beijing, 100048, China}
\date{\today }

\begin{abstract}

The spontaneously broken of translational symmetry is usually due to the competition between local interactions and long-range interactions. However, in this paper, we show how a crystalline order can be generated by the competition between local interaction and long-range correlation by frustration. Here we propose a positive hopping Bose Hubbard model on triangle lattices with a pair creation term which comes from frustrated linked cavity arrays with degenerate quantum gases in them.  We find by increasing the strength of pair creation term against local interaction strength, two kinds of density wave ordered superradiant photonic fluid phase can be realized and a first order transition between two different density wave ordered states is found. This proposal shows us a new way to produce coherent  ``solid" phase without the help of long range interactions.

\end{abstract}

\maketitle

\emph{Introduction.} 
Recently, superradiance in a cavity has been realized experimentally  \cite{Black, Esslinger1, Esslinger2} due to the technique advances developed in strong atom-light coupling in cavity systems\cite{Kimble, Esslinger0, Reichel}. Superradiance is a coherent state of strongly interacting atoms and light, which is originally proposed by Dicke in 1954\cite{Dicke}. The superradiance transition in a single mode cavity spontaneously break a ${\mathbb Z}_2$ symmetry, where the phase of cavity field can only be 0 or $\pi$ after condensation. The phase lock between cavity field and pumping field, together with the even odd lattice site switch in density pattern of atoms was observed in a following experiment\cite{Esslinger2}. One can find the 0 and $\pi$ phase of the cavity field is quite alike a ferromagnetic state with spin up and down. The only difference between a condensed cavity field and the spin is the magnitude of the cavity field is obtained by condensation, so it has self adjusted ability while the magnitude of spin is fixed without fluctuation.  

When classical spins are put on three sites with anti-ferromagnetic coupling, strong frustration is generated and the energy of different classical spin configurations are degenerate. If one extend the three sites into a triangle lattice, there are infinite many degenerate classical spin configurations and the true ground state is chosen by quantum fluctuations\cite{Wannier, Houtappel, Sondhi00, Sondhi01, Moessner03}. Similarly, we could add positive hopping between cavities to generate frustration between three superradiant condensates. However, in a superradiant condensate, the phase of the condensate is not exactly locked at $0$ and $\pi$ and the amplitude of condensate is not fixed as well. Comparing with spin model, the phase fluctuations and the magnitude fluctuations of the condensate loose the constraint of frustration a little bit. These photon density fluctuations and phase fluctuations are suppressed by atom-light coupling strength. For small atom-light coupling strength, the phase of condensate could be any value from $0$ to $2\pi$, which is more similar to U(1) symmetry case. For extremely large atom-light coupling strength, the phase of the condensate is focused on $0$ or $\pi$, which is similar to Ising spin case (${\mathbb Z}_2$). For this reason, by tuning the interaction strength of atom-light coupling, we have a symmetry crossover process from U(1) to ${\mathbb Z}_2$.

In this paper, we find the symmetry crossover can give rise to a density wave superradiance as the ground state of photons in frustrated triangle cavity arrays. To illustrate this point, an effective positive-hopping Bose-Hubbard model with pair generation term is introduced as our starting point. 

\ba
\hat{H}&=&J\sum_{\langle ij\rangle} a_{\bf i}^\dag a_{\bf j}-\mu\sum_{\bf i}\hat{n}_{\bf i}\nonumber\\
&&+\frac{U}{2}\sum_{\bf i}\hat{n}_{\bf i}(\hat{n}_{\bf i}-1)-\chi\sum_{\bf i}(a_{\bf i}+a^\dag_{\bf i})^2,\label{Hamil}
\ea
where the system is put on a two-dimensional triangle lattice. $J>0$ is the hopping strength, $\mu$ is the chemical potential, $U$ is the onsite interaction energy and $\chi$ is an induced interaction strength. $\hat{a}_{\bf i}$ is boson annihilation operator on cavity site ${\bf i}$. $\hat{n}_{\bf i}=a^\dag_{\bf i}a_{\bf i}$. Without the last term, this is a standard frustrated Bose Hubbard model, whose ground state is 120 degree ordered superfluid state\cite{Diep04,Cirac08}. In $U\rightarrow\infty$ limit, the frustrated Bose Hubbard model can be mapped into XY spin model, whose ground state is 120 degree magnetic order\cite{Diep04}. However, when $\chi$ term is added, photon condensate favors $0$ or $\pi$ phase, $U(1)$ symmetry is explicitly broken to ${\mathbb Z}_2$, the frustration becomes Ising-like. But for a bosonic system, the condensate on every site can be adjusted self-consistently. This is quite different from spin systems. Then we find photon condensate could develop a density wave pattern to avoid strong frustration. Hence, a density wave superradiance is generated by competition between local interaction and correlation induced by frustration during a symmetry crossover from U(1) to ${\mathbb Z}_2$.

In the following, we will first discuss how to setup our system. Then we construct our solution from a three site problem to a cluster solution, and then we extend the solution to a lattice system. Finally, we give a mean field study of the frustrated lattice system for ground state to justify our claim.

\emph{Experimental Setup} Recently, superradiance followed by strong atom-light coupling has been realized in experiments. In the experiment, the coupling between the cavity field and the atoms can be given in terms of
\ba
\hat{H}_{\rm int}=(\hat{a}+\hat{a}^\dag)\hat{\Theta}+\hat{a}^\dag\hat{a}\hat{\cal B},
\ea
where $\hat{\Theta}=\int d{\bf r} \hat{n}_{\rm at}({\bf r})\eta({\bf r})$ and ${\cal B}=\int d{\bf r} \hat{n}_{\rm at}({\bf r})U_0({\bf r})$ are two density orders of the atomic gases\cite{Esslinger1}. $\hat{n}_{\rm at}({\bf r})$ is atomic density operator, $\eta({\bf r})$ and $U_0({\bf r})$ are two modes functions. Terms like $-\chi(a^\dag+a)^2$ and interaction terms like $(U/2)\hat{n}(\hat{n}-1)$ can be generated by $\langle \hat{\Theta}\hat{\Theta}\rangle (\hat{a}+\hat{a}^\dag)^2$ and $\langle \hat{\cal B}\hat{\cal B}\rangle \hat{a}^\dag \hat{a}^\dag \hat{a}\hat{a}$, where $\langle\cdot\rangle$ are ensemble average. The strength of the $\chi$ and $U$ can then be tuned by the strength of $\eta({\bf r})$ and $U_0({\bf r})$ independently. Other terms are neglected for simplicity. By this setup, we can access the onsite terms in Eq.(\ref{Hamil}).
\begin{figure}[t]
\includegraphics[width=7.5cm]{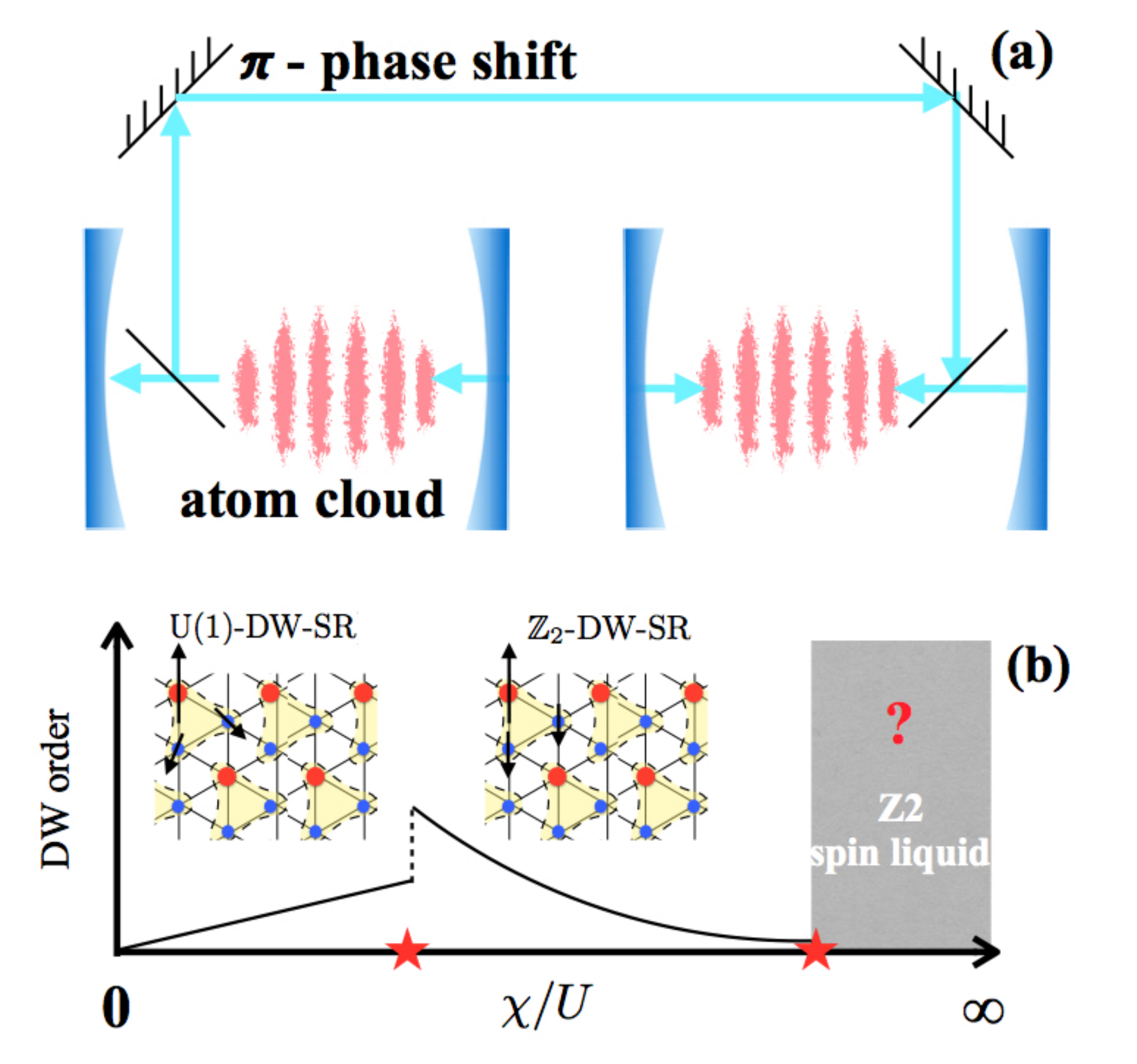}
\caption{In (a), we give a experimental set up to realize a hopping between two cavities with a $\pi$ phase shift. In (b), we show a typical phase diagram for varying $\chi/U$ at fixed $J/U$, and $\mu/U$. When $\chi/U$ becomes larger, the system's ground state. The vertical axis gives density wave order }\label{Setup}
\end{figure}

On the other hand, we design a scheme for cavity photons to tunnel between cavities with a phase shift. In Fig. \ref{Setup}(a), we set a half transmissive half reflective mirror in one cavity so that cavity photon can leave. There are other two mirrors which can produce a $\pi$ phase shift during the photon propagating by a half-wave loss mechanism. Finally, the photon enters another cavity by another half transmissive half reflective mirror. Through this setup we can have a positive hopping between two cavities (J term), thus it is possible to realize Eq.(\ref{Hamil}). 

Finally, to simplify our problem, we neglect the cavity decay rate $\kappa$ through out this paper.

\emph{A three sites problem.} Here we try to construct our solution from few sites to many. Hence we start from a three site problem. Neglecting $J$, we find the cavity field $\hat{a}$ will condense into a superradiant state when $\chi$ is large enough to overcome the red detune of the cavity. We can find the condensate of $\langle\hat{a}\rangle$ has two equivalent phases to choose, $0$ and $\pi$. Therefore when a anti-ferromagnetic coupling $J>0$ is turned on between adjacent cavities, $\alpha_{\bf i}=\langle \hat{a}_{\bf i}\rangle$ favors anti-parallel configuration. Meanwhile, unlike the spin model which is completely a non-linear sigma model where the amplitude mode is infinite heavy, the magnitude of onsite condensation is variable in the present situation. 

Here we apply a mean field theory with independent local order parameters $\alpha_{\bf i}=\langle \hat{a}_{\bf i}\rangle$ to Eq.(\ref{Hamil}) on three sites. Assuming the ground state is $|\Omega\rangle=|\alpha_1\rangle\otimes|\alpha_2\rangle\otimes|\alpha_3\rangle$, where $|\alpha_i\rangle$ is coherent state on site ${\bf i}$. The self-consistent equations can be obtained by minimizing the ground state energy $E(\alpha_1,\alpha_2,\alpha_3)=\langle\Omega|\hat{H}|\Omega\rangle$ as 
\ba
-\left(\mu\!+\!\chi\!+\!\frac{U}{2}\right)\alpha_a\!+\!U|\alpha_a|^2\alpha_a\!+\!J(\alpha_b\!+\!\alpha_c)=2\chi{\rm Re}(\alpha_a),\label{MFT}
\ea
where $(a,b,c)=(1,2,3)$, $(2,3,1)$ and $(3,1,2)$. ${\rm Re}(\#) ({\rm Im}(\#))$ takes the real part (imaginary part) of $\#$. Since the equations are complex, there are indeed 6 equations. From here on we will always take U=1 as our unit energy, other energies should be understood as $\mu/U$, $\chi/U$ and $J/U$.
\begin{figure}[t]
\includegraphics[width=8.5cm]{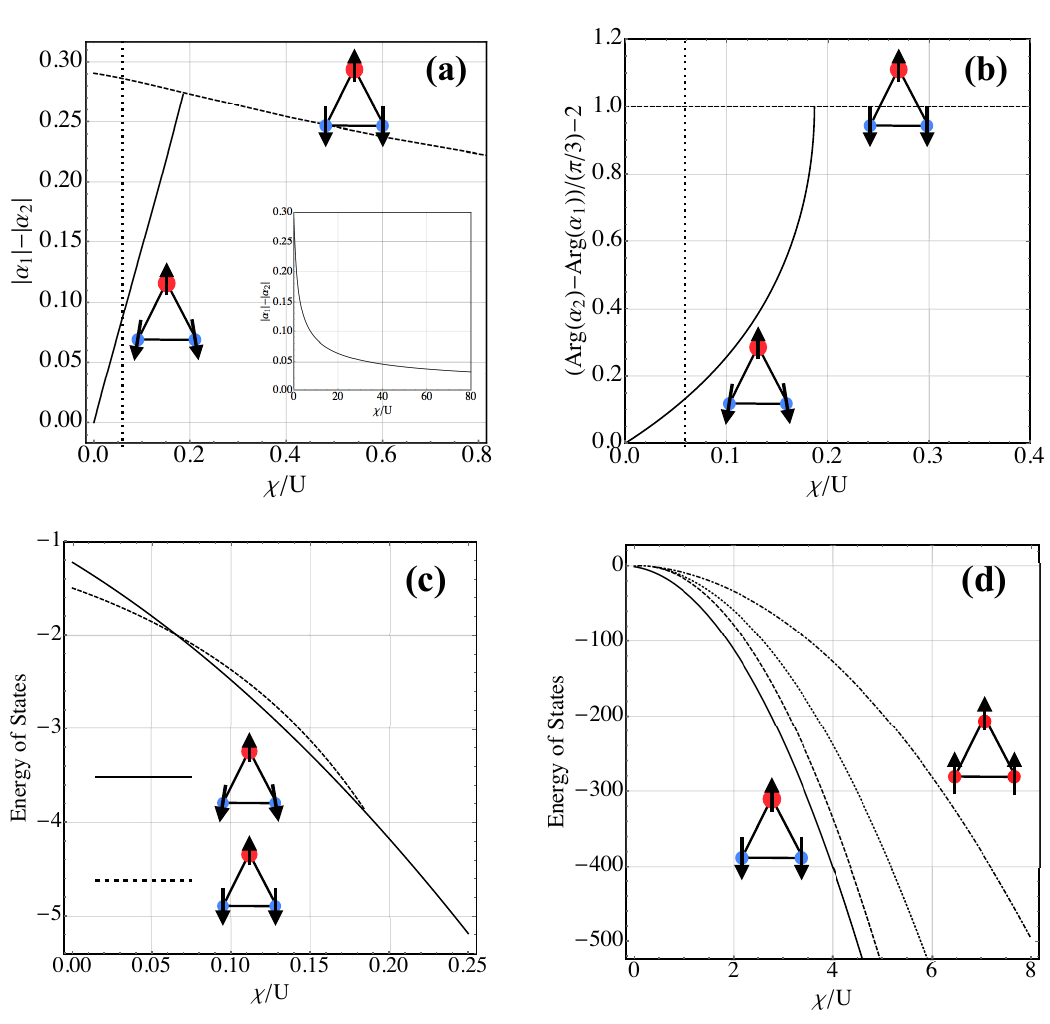}
\caption{In (a), we give the density order $|\alpha_1|-|\alpha_2|$ for the first solution U(1) DW ordered state and the second solution ${\mathbb Z}_2$ DW ordered state in solid line and dashed line respectively. (b) The phase difference between two sites in U(1) DW ordered state and ${\mathbb Z}_2$ DW ordered state. In (c), we show the energy of U(1) DW ordered state and ${\mathbb Z}_2$ DW ordered state. There is a level crossing before U(1) DW ordered state develop 180 degree between $\alpha_1$ and $\alpha_2$'s phase, hence a first order transition is expected. In (d), we compare the energy of all four solutions for Eq. (\ref{S1}) and Eq. (\ref{S2}). The lowest stable state is the ${\mathbb Z}_2$ DW ordered state.}\label{ThreeSite}
\end{figure}

There are two possible solutions for these equations. The first solution requires ${\rm Im}\alpha_1=0$, ${\rm Im}\alpha_2=-{\rm Im}\alpha_3\neq0$. we have $|\alpha_1|^2=\mu+1/2+3\chi+J/(1-(\chi/J))$, $|\alpha_2|^2=|\alpha_3|^2=\mu+1/2+\chi+J$ and ${\rm Re}(\alpha_2)=J\alpha_1/(2\chi-2J)$. The density order is
\ba
|\alpha_1|^2-|\alpha_2|^2=2\chi+\chi/(1-(\chi/J)).
\ea
This solution requires $|{\rm Re}(\alpha_2)|\leq|\alpha_2|$. At the point $|{\rm Re}(\alpha_2)|=|\alpha_2|$, $\alpha_2$ becomes real.
In this solution, we stress two points. 1) As long as $\chi\neq0$, there is a density imbalance between different cavities. The condensate arranged a density order to reconcile with frustration and the density order is proportional to $\chi$.  2) The relative phase between different sites is not 180 degree, more like 120 degree order. For this reason we call this state a U(1) density wave superradiant state (U(1)-DW-SR). 

In another possible solution, all the imaginary parts are order parameters are zero, that is ${\rm Im}(\alpha_{a,b,c})=0$. The phases of every two neighboring sites are either parallel or anti-parallel. As the equations for $\alpha_2$ and $\alpha_3$ are symmetric, we assume $\alpha_2=\alpha_3$. Then we have
\ba
(\alpha_1^2-\mu-1/2-3\chi)\alpha_1&=&-2J\alpha_2\label{S1}\\
(\alpha_2^2-\mu-1/2-3\chi+J)\alpha_2&=&-J\alpha_1\label{S2}
\ea
These two equations can be solved analytically and there are four solutions in total. These four solutions' energy are shown in Fig.\ref{ThreeSite}(d), and we find the lowest energy state takes one ``spin up" large condensate, two small ``spin down" condensates configuration. The density difference $\Delta_{\rm CDW}=|\alpha_1|-|\alpha_2|$ is shown in Fig.\ref{ThreeSite}(a) as the dashed line. The density difference vanishes when $\chi\rightarrow\infty$ and this is shown in the inset figure of Fig.\ref{ThreeSite}(a). The phase difference $({\rm Arg}(\alpha_1)-{\rm Arg}(\alpha_2))/\pi$ is shown in Fig.\ref{ThreeSite}(b). From all these features we find the second solution is a density wave ordered ``antiferromagnetic" state. Here we call it ${\mathbb Z}_2$ density wave superradiant state (${\mathbb Z}_2$-DW-SR). 

To summarize, the ground state is either U(1)-DW-SR or ${\mathbb Z}_2$-DW-SR. When $\chi$ is small, a U(1)-DW-SR is the true ground state and when $\chi$ becomes large, a ${\mathbb Z}_2$-DW-SR is the real ground state. The density order changes non-monotonously against $\chi$. When U(1) symmetry is weakly broken by $\chi$, a density order is generated to reconcile the frustration. However, density wave order is also suppressed by large $\chi/U$. That is because the density fluctuations of $|\alpha_a|^2$ are greatly suppressed by exact ${\mathbb Z}_2$ symmetry imposed by $\chi$ term. Large $\chi$ term fix $\alpha$. For infinite large $\chi$, we find the density order disappears, three configurations of two "spin up" and two "spin down" are degenerate, which becomes a frustrated Ising model on triangle lattices. 
\begin{figure}[b]
\includegraphics[width=7.5cm]{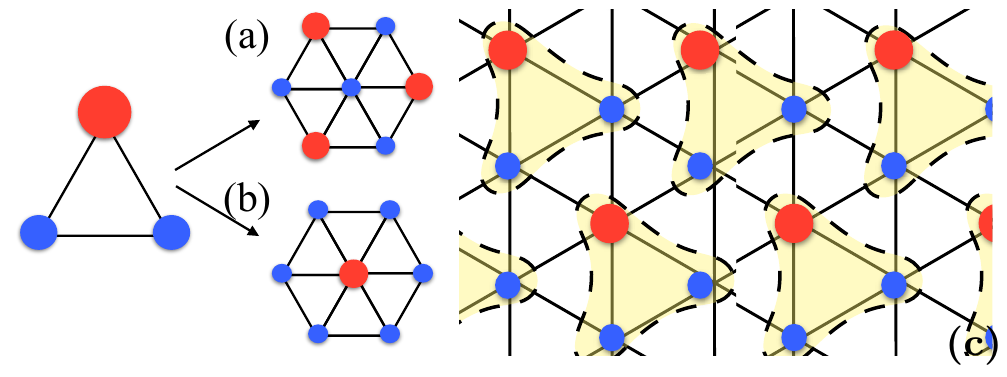}
\caption{When we increase the site number to be 7, the possible solutions are given in (a) and (b), where the size of the spot represents the local density and the red and blue color means two nearly opposite phases. (b) is lowest energy configuration, (a) is a possible saddle point solution. Combining these two solutions, a lattice solution can be constructed by adding site one by one. In (c), we show the ground state configuration suggested by the few site solution in triangle lattices. A super unit cell can be found, which is circled by dashed lines in (c).}
\label{Cluster}
\end{figure}

\emph{Few sites problem. } From the three sites problem, we learn the optimized configuration for three sites is one large condensate with two small condensates. One criteria can be established that between large-small bond, two condensates form an anti-ferromagnetic pair, and between small-small bond, two condensates form a ferromagnetic pair. By extending this rule to larger clusters, we find there could be two stable 7 sites configurations as is shown in Fig. \ref{Cluster}(a) and (b). A lattice structure can be generated following this rule. 

Here we assume a coherent product state $|\Omega\rangle=\prod_{\bf i\in\Lambda}|\alpha_{\bf i}\rangle$ as the ground state variational wave function. By minimizing ground state energy $E_\Omega=\langle\Omega|\hat{H}|\Omega\rangle$ as a function of $\alpha_{{\bf i}=1,2,\cdots,7}$, we find configuration Fig. \ref{Cluster}(b) minimize $E_\Omega$. The configuration in Fig.\ref{Cluster}(a) is also a steady state solution. These two configurations both satisfy the rules we propose above. By applying this rule we find a periodic structure in triangle lattice where two sub-lattices are formed. The large condensate sites form a ferromagnetic triangle lattice and the small condensate sites form another ferromagnetic hexagon lattice. The phases between these two sub-lattices are opposite.

\emph{Mean Field Theory.} Now we begin to analyze Eq.(\ref{Hamil}) on a two-dimensional lattice system by a mean field theory. With previous knowledge on the cluster solutions, we assume the solution has a super unit cell including three sites in it. In Fig. \ref{Cluster}(c), we give the configuration of the lattice structure and the unit cell. One can see every three sites can be taken as a super unit cell. Here we introduce three order parameters in one super unit cell as $\alpha_{{\bf i}a}$ where $a=1,2,3$ are inner super unit cell index and ${\bf i}$ is the super unit cell position. 

First, if we neglect the spatial fluctuations and assume $\alpha_{{\bf i}a}=\alpha_a$, we find the ground state energy density is,
\ba
3E_{G}/N_\Lambda&=&3J\sum_{a\neq b}\alpha_a^*\alpha_b-\left(\mu+\frac{U}{2}\right)\sum_a|\alpha_a|^2\nonumber\\
&&+\frac{U}{2}\sum_a|\alpha_a|^4-\chi\sum_a(\alpha_a+\alpha_a^*)^2.
\ea
To minimize the ground state energy, we have
\ba
\!-\!\left(\mu\!+\!\chi\!+\!\frac{U}{2}\right)\alpha_a\!+\!U|\alpha_a|^2\alpha_a\!+\!3J(\alpha_b\!+\!\alpha_c)\!=\!2\chi{\rm Re}(\alpha_a),\label{MF}
\ea
where $(a,b,c)=(1,2,3)$, $(2,3,1)$ and $(3,1,2)$. One can find this equation is the same as Eq. (\ref{MFT}) when $3J$ is replaced by $J$. Therefore previous solutions for three site problem can be translated into phase diagram of the system. As we learn from the three site problem, there are two kind s of solutions. The U(1) DW ordered state represents a density wave ordered superfluid phase with phase not exactly $180^{\rm o}$ (U(1)-DW-SR), and the ${\mathbb Z}_2$ DW ordered state represents a density wave ordered superfluid phase with phase difference between large condensate site and small condensate site being $180^{\rm o}$. According to Fig. 4(c), the mean field transition between these U(1)-DW-SR and ${\mathbb Z}_2$-DW-SR is a first order transition with a jump in both the superfluid order and the density order. We here present the phase diagram based on Eq. (\ref{MF}) and comparison of ground state energy. A density order is discovered when the original hamiltonian is subject to a translational invariance and all interaction is local. A translational invariance is broken by a internal symmetry breaking. Interestingly, this density order is non-monotonously dependent on $\chi/U$. When $\chi/U$ is small, arbitrary small $\chi$ brings the system a density order. When $\chi/U$ is large, the symmetry of the system is fixed to ${\mathbb Z}_2$ and the local density fluctuations are suppressed. As a result, density order is weakened for large $\chi/U$.

\begin{figure}[ht]
\includegraphics[width=8.5cm]{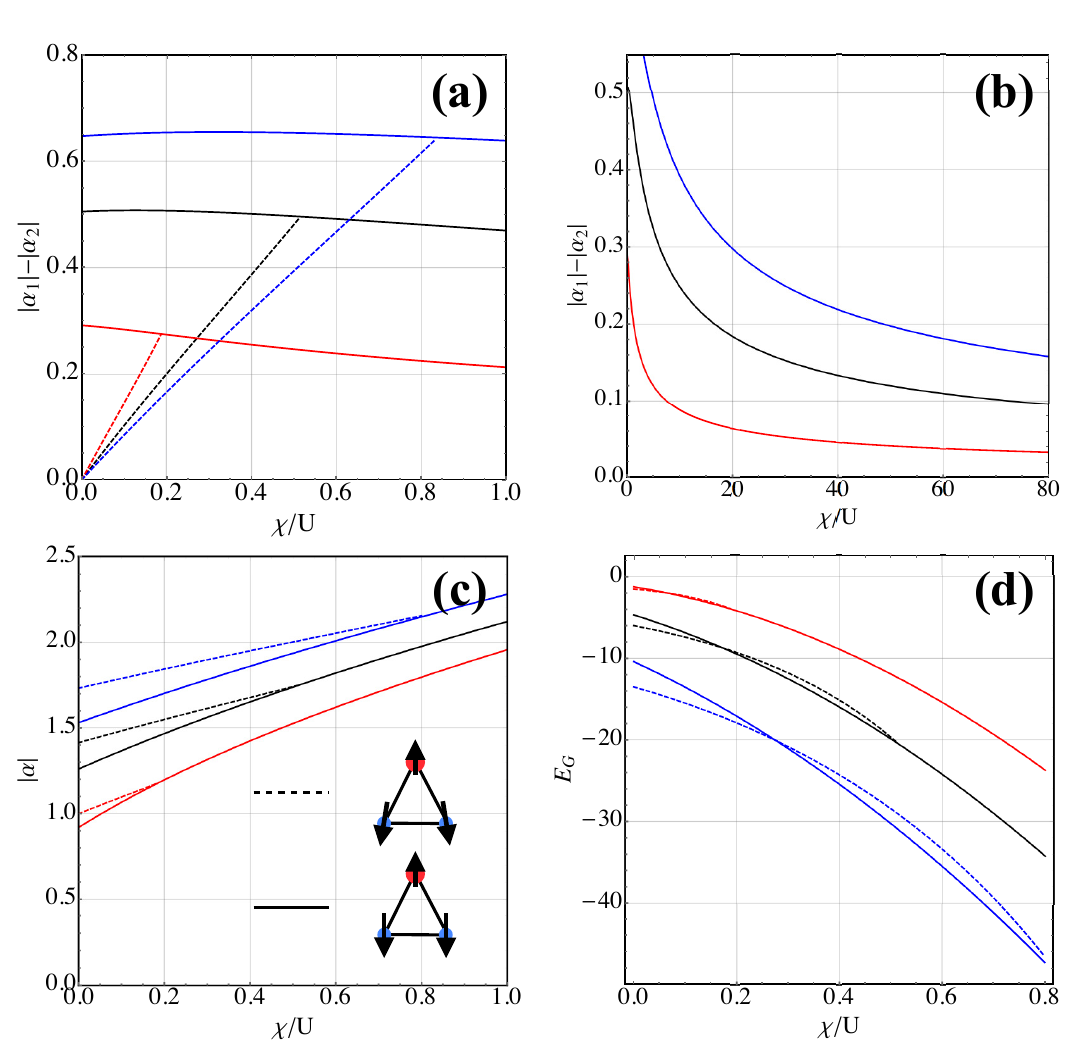}
\caption{In (a), (b), we show the density wave order against $\chi/U$ for fixed $J/U$ and $\mu/U=0$. $3J/U=$2.5, 1.5, and 0.5 are shown as blue lines, black lines and red lines. The dashed lines are for density wave A phase, and the solid lines are for density wave B phase. In (c) we show the average cavity field strength $|\alpha|=\sum_{{\bf i}a}\langle \hat{a}_{{\bf i}a}\rangle/N_\Lambda$, where $N_\Lambda$ is the total lattice sites number. In (d), we show the energy curves for density wave A phase and density wave B phase in dashed lines and solid lines respectively. A first order transition between these two phases are implied by the energy crossing points. }
\end{figure}

There are two factors that are beyond a mean field theory. Here we analyze the impact of these two factors. The first factor is the spatial phase fluctuations of $\alpha_{{\bf i}a}$. We find there can be a uniform phase fluctuation in a super unit cell, $\alpha_{{\bf i}a}=\alpha_a e^{i\varphi_{\bf i}}$. One can find for a density operator, the local phase is always cancelled. Hence the density wave order is irrelevant to this phase fluctuation.

The second factor is quantum fluctuations between nearly degenerate configurations. As we learn in a three sites problem, when $\chi/U$ is large, the phase fluctuations of local condensate is greatly suppressed, and the density wave order is also suppressed. For a large $\chi/U$, the energy of different configurations in super unit cell is nearly degenerate. Hence it is more similar to frustrated Ising model. For a finite large $\chi/U$, we start from a homogeneous configuration and ask how much a density order can lower the energy and how much a ${\mathbb Z}_2$ spin liquid could lower the energy. Both two schemes lowers the energy from the highly degenerate starting point, therefore the true ground state should be the one lower more energy from this configuration. As the density order is vanishing in $\chi/U\rightarrow\infty$ limit, so the energy gain by density order goes to zero in this limit. On the other hand, the contribution from quantum fluctuation is nonzero in the same limit. Therefore we expect a direct phase transition between these two states at finite $\chi/U$. However, to verify the existence of ${\mathbb Z}_2$ spin liquid and the phase transition requires methods which can properly count quantum fluctuations. This is beyond the scope of the present paper. We will leave this interesting phenomenon for the future study. 

\emph{Conclusion}. In this paper, we propose a positive hopping Bose-Hubbard model with a pair-generation term on a triangle lattice. Through a mean field study we extend our few site solutions to a lattice solution. We find the ground state of this model shows both off-diagonal long range order and crystalline diagonal long range order. There are two density wave coherent photonic fluid state phase transition to each other by a first order transition. The density wave order is non-monotonously dependent on the pair-generation term, which is small for both small and large pair-generation term. Thus we give an example to generate density by competition between local interactions and long range correlation induced by frustration where original model have only local interactions and is translational invariant.

\emph {Acknowledgement}.
This work is supported by NSFC under Grant No. 11604225 and No. 11734010, Beijing Natural Science Foundation (Z180013), Foundation of Beijing Education Committees under Grant No. KM201710028004. We thank Pengfei Zhang and Hui Zhai for discussions.


\end{document}